# Impact of Road Infrastructure and Traffic Scenarios on E-scooterists' Riding and Gaze Behavior

**Dong Chen[1], Arman Hosseini[2], Arik Smith[3], Zeyang Zheng[4], David Xiang[5], Arsalan Heydarian[6], Omidreza Shoghli[7], and Brad Campbell[8]**

[1] Agriculture and Biomedical Engineering, Mississippi State University, Mississippi State, MS 39762, USA; e-mail: dchen@abe.msstate.edu

[2,3,4,5] Systems Engineering and Computer Science, University of Virginia, Charlottesville, VA, 22903, USA; e-mail: ufy5tc@virginia.edu , qgn5zs@virginia.edu , yuq8cp@virginia.edu , mwd4wm@virginia.edu

[6] Civil and Environmental Engineering, University of Virginia, Charlottesville, VA, 22903, USA; e-mail: heydarian@virginia.edu

[7] Civil Engineering Technology and Construction Management, University of North Carolina at Charlotte, Charlotte, NC 28223, USA; e-mail: oshoghli@charlotte.edu

[8] Link Lab & Computer Science, University of Virginia, Charlottesville, VA, 22903, USA; e-mail: bardjc@virginia.edu

## ABSTRACT

The growing adoption of e-scooters has raised significant safety concerns, particularly due to a surge in injuries and fatalities. This study explores the relationship between road infrastructure, traffic scenarios, and e-scooterists' riding and gaze behaviors to improve road safety and user experience. A naturalistic study was conducted using instrumented e-scooters, capturing gaze patterns, fixation metrics, and head movement data across various road layouts and traffic scenarios. Key findings reveal that bike lanes offer a stable environment with reduced horizontal head movement and focused attention on the road, while shared roads and sidewalks lead to more dispersed gaze and increased head movement, indicating higher uncertainty and complexity. Interactions with other road users—such as navigating intersections, passing buses, riding near cars, and descending on downhill paths—demand greater cognitive load. Intersections re- quire heightened visual focus and spatial awareness, reflected in increased horizontal eye and head movements. Interactions with vehicles prioritize visual scanning over head movement to maintain stability and avoid collisions, while high-speed and downhill riding demand focused attention on obstacles and the road surface. The results provide insights into e-scooter riders' behavior and physiological response analysis, paving the way for safer riding experiences and improved understanding of their needs.

# INTRODUCTION

As e-scooters gain popularity as a convenient mode of transportation, their integration into urban settings has raised critical safety concerns, highlighting the need for effective strategies to address these challenges. Between 2017 and 2021, injuries related to micromobility vehicles experienced a staggering 127% increase, reaching 77,200 reported cases. Among these, e-scooters accounted for the most pronounced rise in both injuries and fatalities, underscoring the urgent need for enhanced safety measures and informed urban planning to accommodate this growing mode of transport. (Tark 2022). Specifically, according to the U.S. Consumer Product Safety Commission (CPSC), e-scooters have seen a continuous year-over-year increase in injuries, with a 22% rise in 2022 compared to 2021 (CPSC). A comprehensive study by the Centers for Disease Control and Prevention (CDC) (Hayden and Spillar 2019) found that 20 out of every 100,000 e-scooter trips resulted in injuries. Half of these injuries are head injuries, with a significant 15% related to severe traumatic brain injuries. Even more alarmingly, e-scooter-related injuries among children have risen noticeably, with hospitalizations increasing from 4.2% in 2011 to 12.9% in 2020 (Hayward 2022, Krištopaityte˙ 2022). Given these alarming trends, the urgent need for innovative research to bridge the safety knowledge gap becomes evident, paving the way for safer e-scooter integration.

Previous research has primarily used emergency room records and media reports to gather crash information and conduct safety studies (Badeau et al. 2019 , Yang et al. 2020). For instance, the study by (Badeau et al. 2019) analyzed treatment records from two emergency departments, revealing that 44% of e-scooter incidents took place on sidewalks. Similarly, (Yang et al. 2020) utilized a comprehensive collection of news reports detailing 169 e-scooter-involved crashes across the US between 2017 and 2019, finding that the majority of these incidents occurred on streets, intersections, and sidewalks. Additionally, survey and observational studies have been pivotal in this realm (Zou et al. 2020). For example, targeted research in Washington, DC, USA, involving roadside observations of 2,004 e-scooters, found that riders more commonly used sidewalks (60%) rather than roads (40%) in areas devoid of bike lanes. These observations stress the importance of exploring innovative strategies to comprehend the behavioral patterns and psycho-physiological responses of e-scooter users in various roadway conditions and infrastructures to enhance safety measures (Cicchino et al. 2023).

Despite these valuable insights, the aforementioned studies have not adequately captured the critical situational interactions between riders and the other road users that precipitate accidents. This gap underscores the critical need for naturalistic studies that observe and record behaviors and responses in the complex, uncontrolled conditions of real-world settings (Janikian et al. 2024). Such studies are uniquely positioned to reveal the nuanced dynamics of e-scooter usage, offering insights into how different infrastructural elements and traffic scenarios directly impact rider safety. Nevertheless, focused research on e-scooter safety through naturalistic methods remains relatively scarce (Janikian et al. 2024). Notably, (Cano-Moreno et al. 2022) and (Ma et al. 2023) delved into the naturalistic studies of risks associated with e-scooter vibrations, which can impact riding comfort due to varying road surfaces and wheel sizes. Moreover, (Ma et al. 2021) analyzed e-scooter interactions with urban infrastructure, revealing that vibrations, speed variations, and environmental constraints elevate safety risks, emphasizing the necessity for research to enhance infrastructure and policies. Similarly, (Pashkevich et al. 2022) investigated the visual attention of pedestrians, cyclists, and e- scooter users through mobile eye tracking, showing that while fixation rates per minute were comparable across these groups, the distribution of their visual attention varied significantly. However, this research did not cover diverse traffic conditions, and its analysis was confined to fixation metrics, omitting critical metrics like gaze entropy and variability (Guo et al. 2024).

This paper explores the interplay between e-scooter users' riding and gaze behavior across various traffic scenarios and infrastructures, utilizing naturalistic data to understand their impact on road safety and user experience. It examines road layouts, including sidewalks, pedestrian trails, and roads with or without bike lanes, as well as traffic scenarios such as passing pedestrians and interacting with cars at intersections. Using metrics like head movement variability, gaze density heat maps, road center focus, gaze entropies, and fixation patterns, the study provides a detailed analysis of riders' visual and cognitive engagement.

**METHODOLOGY**

**Experiment setup.** The naturalistic study was conducted using a Ninebot MAX KickScooter, equipped with a 350W motor capable of reaching speeds up to 18.6 mph and a range of 25 miles, supporting riders up to 220 lbs. Additionally, the SpeedTracker app further enhances data collection, recording vital travel information such as GPS, speed, elevation, and timestamps at one-second intervals (see Figure 1). Instrumentation also included Tobi Pro Glasses 3, which captured a wealth of data on gaze behavior, including gaze patterns, eye movements, saccades, and visual fixations. This device, equipped with a 3-axis gyroscope and accelerometer, recorded continuous forward-facing video at a resolution of 1920 × 1080 pixels with a 90-degree field of view at 25 Hz. Additionally, a Samsung Galaxy smartwatch was used to record heart rate variations. Although this study primarily focuses on the visual data, a detailed analysis of heart rate variations across different scenarios is planned for future research.

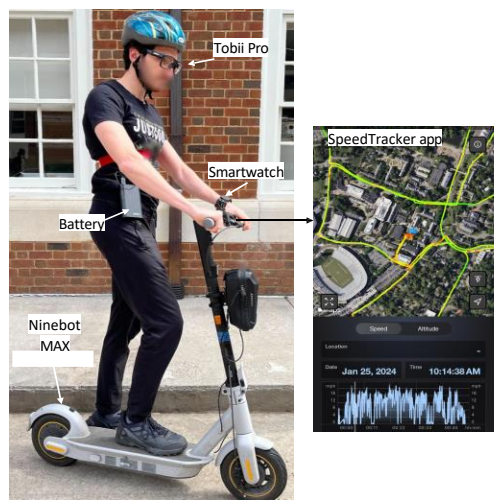

**Fig. 1. E-Scooter instrumentation for naturalistic study.**

**Experiment procedure & Data cleaning.** The naturalistic e-scooter rides were conducted along diverse routes around the University of Virginia (UVA) and downtown Charlottesville, VA, USA, to explore various road structures and traffic scenarios. To ensure a comprehensive evaluation, each experiment featured different routes. All rides were conducted during daylight hours, lasting approximately one hour each, under sunny or cloudy weather conditions, with no trials conducted in the rain. Before each ride, pre-ride safety inspections and meticulous equipment calibrations were performed to ensure the reliability of all data collection tools.

Five participants (4 male, 1 female), all members of the research lab, participated in the study, contributing approximately 4 hours of data each, totaling around 20 hours of riding. The group included three graduate students aged 22–28 and two undergraduate students aged 18–22. Among them, three participants were regular and experienced e-scooter riders, while the other two were infrequent users who had previously only tried shared e-scooters. Each trip yielded comprehensive data, including speed, GPS coordinates, elevation, and gaze metrics, all synchronized with the frame rate of the gaze data for precise analysis. Data cleaning involved filtering noise from the GPS and gaze data, addressing missing values, and resolving inconsistencies to ensure a high-quality dataset. Traffic scenarios, including interactions with other road users and riding on various infrastructures, were categorized using video recordings captured by the camera integrated into the smart glasses. The identification of interactions with other road users was performed partly manually by the research team and partly through the object detection dataset for e-scooters developed by (Chen et al. 2024). Detailed methodology and data processing techniques are available in our open-source repository (*https://github.com/DongChen06/Escooter_ITSC*).

**Gaze and head movement analysis.** Eye gaze patterns have been widely utilized in research to infer various human cognitive and emotional states. Metrics such as fixation duration, direction-specific gaze variability, and gaze entropies have proven to be valuable indicators, showing strong correlations with cognitive workload, stress levels, and emotional states (Shiferaw et al. 2019 , Guo et al. 2022). Fixations, occurring when the eyes pause scanning and maintain central foveal vision on a specific point, are among the most commonly analyzed features in eye-tracking studies to infer cognitive processes and states. The Tobii Pro Lab software, used with Tobii glasses, employs a built-in algorithm that classifies any gaze lasting 0.02 seconds or more as a fixation. Fixation duration is associated with the cognitive processing time required for the observed target, reflecting efforts to structure these significant eye movements into meaningful interpretations of attention, visibility, mental processing, and comprehension (Joseph and Murugesh 2020).

*Stationary Gaze Entropy* (SGE) measures the predictability of fixation locations (Shiferaw et al. 2019), indicating gaze dispersion during a viewing period and quantifying attention distribution as follows:

$$H_{SGE}(x) = -\sum_{i=1}^{n} p_i \log_2 p_i \quad (1)$$

where $H_{SGE}(x)$ represents the entropy value for a given set $x$ (each time bin within each scenario), with $i$ denoting each fixation's location on a 2D plane within $x$, $n$ is the total number of fixations in $x$, and $p_i$ is the proportion of fixations at each state space within $x$. An increase in SGE is linked to greater task difficulty and complexity (Di Stasi et al. 2016).

*Gaze Transition Entropy* (GTE) extends this analysis to the dynamics of gaze shift (Shiferaw et al. 2018), applying conditional entropy to first-order Markov transitions between fixations:

$$H_{GTE}(x) = -\sum_{i=1}^{n} p_i \sum_{j=1}^{n} p(i|j) \log_2 p(i|j), \quad i \neq j. \tag{2}$$

where $p_i$ is the stationary distribution across fixation locations, and $p(i|j)$ represents the transition probability to location $j$ from $i$. This equation captures the variability in gaze transitions, reflecting the complexity of visual attention shifts. Higher GTE is associated with greater top-down influence, which is linked to induced anxiety and situational stress in previous studies (Shiferaw et al. 2019). For the analysis under different conditions, fixation coordinates were discretized by categorizing them into spatial bins of 100 × 100 pixels (Guo et al. 2024).

In addition to gaze analysis, the Tobii Pro Glasses 3 incorporate an IMU sensor that captures head movement and rotation across all three axes—yaw (side-to-side shaking), pitch (up-and-down nodding), and roll (sideways tilting)—measured in degrees per second and sampled at approximately 100 Hz. This head movement data complements eye movement analysis, offering a more comprehensive understanding of the physical responses and orientation of e-scooter riders as they navigate diverse road scenarios.

**RESULTS**

**Speeds.** Figure 2, presents the gaze density heat map across various speed profiles. Visual analysis of the gaze heat map reveals that the low-speed scenario exhibits a more dispersed distribution of gazes compared to the other scenarios. In contrast, the high- speed scenario (>20 mph) demonstrates a higher concentration of gazes towards the center of the viewing area, with the medium-speed scenario (10-20 mph) showing a similar, albeit less concentrated pattern.
Figure 3, depicts a representative round trip on an e-scooter beginning and ending at the University of Virginia (UVA) and passing through the downtown area of Charlottesville. Speed variations along the route are visually depicted using color gradients, with darker hues indicating higher speeds. The figure includes detailed views of two segments characterized by high-speed travel on downhill paths. Notably, these segments lack dedicated bike lanes, compelling e-scooter riders to share the road with vehicles and pedestrians, a factor that significantly escalates safety concerns for all road users.

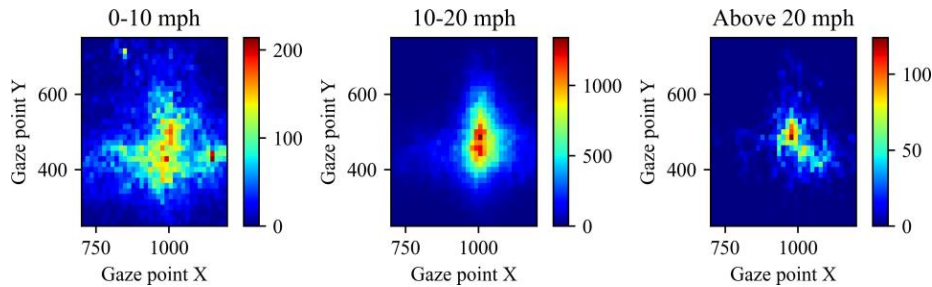

**Fig. 2. Gaze density variation across three speed intervals.**

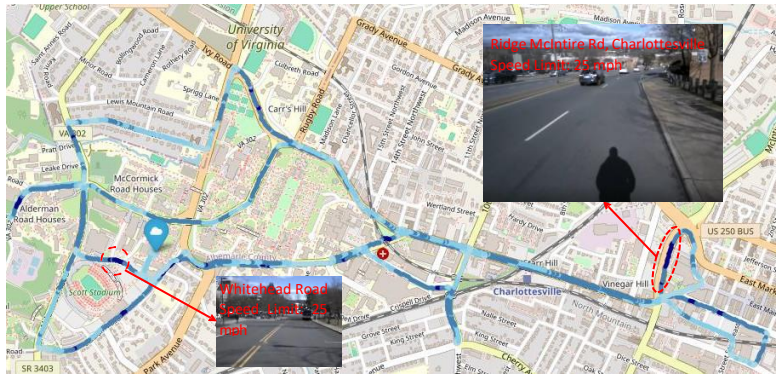

**Fig. 3. Visualization of one representative e-scooter trip.**

**Infrastructures.** This section examines the impact of infrastructure on the visual distribution and head movements of e-scooter riders. Four distinct infrastructure scenarios are analyzed: riding on sidewalks, pedestrian trails (walkways), bike lanes, and shared lanes (Figure 4).

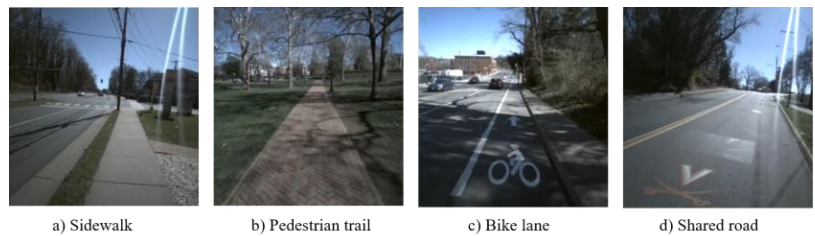

**Fig. 4. Four different infrastructures**

Figure 5, illustrates the gaze density heatmaps of participants during rides across these four infrastructure types. The figure shows that participants exhibited less horizontal gaze movement while riding on bike lanes compared to other infrastructures. However, an increase in vertical gaze variability was observed in the bike lane scenario.

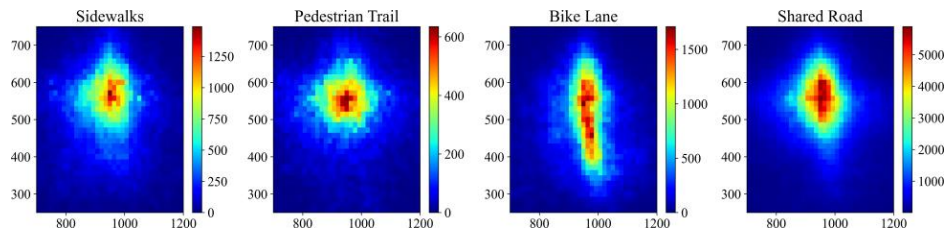

**Fig. 5. Gaze density heatmap for riding on different infrastructures.**

Video analysis revealed that this vertical movement was due to the presence of numerous pinecones along the roadside, prompting riders to pay closer attention to the road surface for safety. In contrast, riding on sidewalks and pedestrian trails showed greater vertical gaze movements, likely attributed to riders observing pedestrians passing by. These patterns highlight how riders adapt their gaze behavior to the specific demands and interactions of each infrastructure type.

Additionally, the vertical head movement of participants, measured by the standard deviation of their head rotation around the Y-axis, is also analyzed. As shown in Figure 6, when e-scooter riders are on the bike lane, they exhibit significantly less horizontal head movement compared to other scenarios. This reduced movement suggests that bike lanes provide a more stable and predictable riding environment, requiring less frequent scanning of the surroundings.

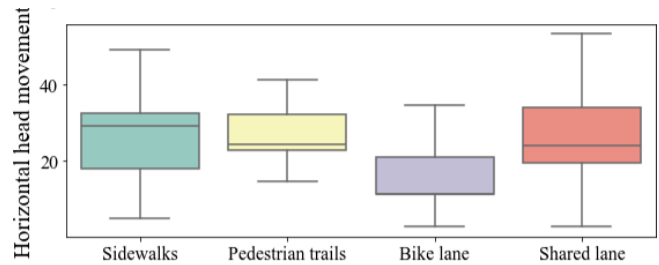

**Fig. 6. Head movement boxplot for riding on different infrastructures**

**Traffic Scenarios.** Another critical factor influencing the safety and stress levels of e-scooter riders is the variety of traffic scenarios they encounter. These scenarios were selected based on survey studies identifying situations perceived as hazardous for either e-scooter riders or other road users interacting with them ([Burt and Ahmed 2023](#)), ([Pourfalatoun et al. 2023](#)). Figure 7, displays an example of each scenario during the ride.

Intersections present significant challenges as they require interactions with both vehicles and pedestrians. Similarly, riding in close proximity to cars or passing by a bus can be intimidating for e-scooter riders, who are among the most vulnerable road users. These scenarios heighten tension and demand increased attention due to the potential risks associated with sharing space with larger vehicles. Two types of interactions with pedestrians, including pedestrians crossing the road and passing pedestrians from side on sidewalks and walkways, are significant scenarios due to the safety concerns they pose for both riders and pedestrians. Additionally, areas with notable downhill segments were studied, as downhill riding is often associated with exceeding safe speed limits. Lastly, two types of transitions were analyzed: moving from a shared road to a crosswalk and transitioning from a bike lane to a shared road. These transitions often involve abrupt changes in the riding environment, requiring riders to adapt their behavior and focus, potentially impacting their safety.

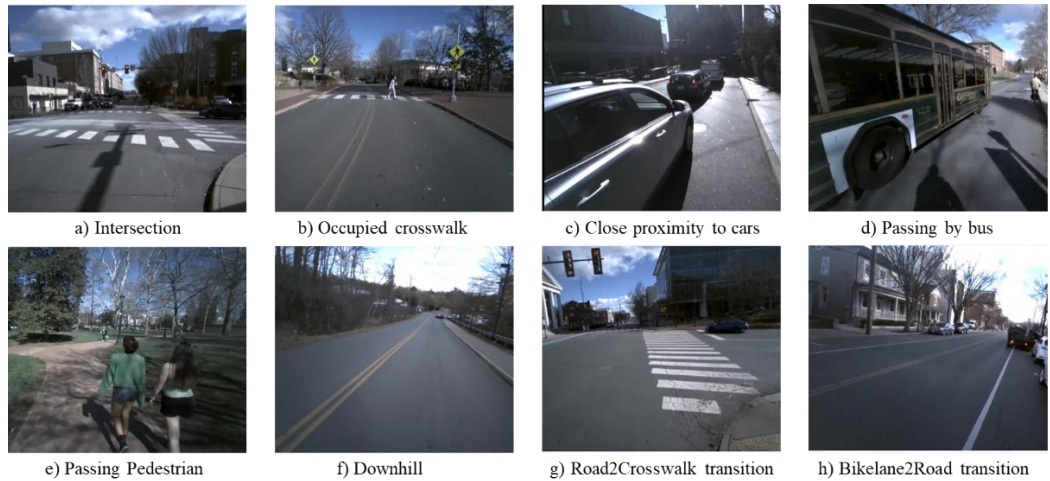

**Fig. 7. Eight traffic scenarios**

Figure 8, depicts the gaze density heatmap across all traffic scenarios, showing distinct visual attention patterns of e-scooter riders. During the "Road to Crosswalk" scenario, the gaze is slightly dispersed but primarily focused around the center, suggesting that the rider is scanning for pedestrians or vehicles while preparing to transition onto the crosswalk. Similarly, in the "Bike Lane to Road" scenario, gaze is distributed along the path, indicating that the rider is planning and navigating the transition from the bike lane to the road while maintaining awareness of the surrounding environment. At "Intersections," gaze is tightly concentrated, reflecting heightened focus on navigating cross-traffic or other infrastructural features. In the "Crossing Pedestrian" scenario, gaze remains predominantly central, highlighting the rider's attention to the pedestrian's movement. The "Downhill" scenario shows the most concentrated gaze pattern, likely due to the need for precise control and heightened awareness of potential obstacles or changes in the higher speed. In "Close Proximity to Cars," gaze is narrowly focused vertically more than horizontally, suggesting that the rider is closely monitoring nearby vehicles and the narrow road he is riding on to avoid collisions. During "Passing Bus," gaze is more scattered horizontally, indicating attention to the bus's movement and its proximity to the rider. Finally, in the "Passing Pedestrian" scenario, gaze shows moderate concentration, likely reflecting a balance of attention between the pedestrian and other road elements.

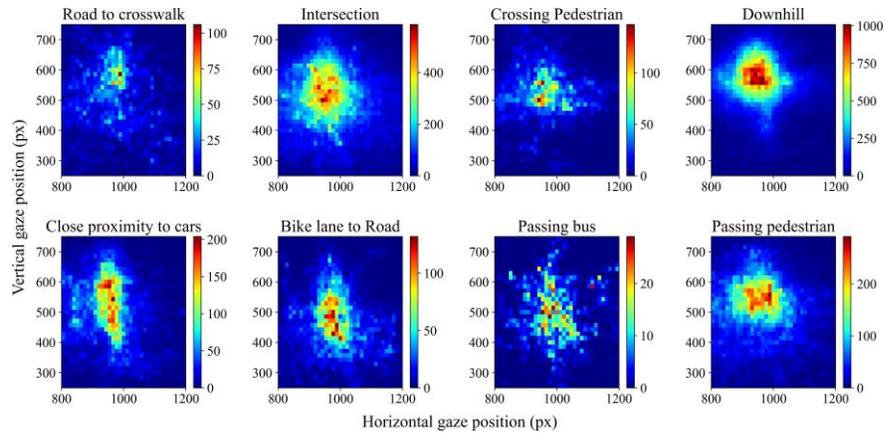

**Fig. 8. Gaze density heat maps for different traffic scenarios.**

Figure 9, illustrates the vertical gaze variability and vertical head movement across various scenarios. The results indicate that scenarios such as transitioning to a crosswalk from a road, moving from a shared road to a bike lane, and navigating intersections exhibit the highest vertical head movements. These scenarios likely demand significant physical adjustments and heightened spatial awareness, contributing to the observed increase in head motion. Interestingly, in situations like being passed by a bus or riding in close proximity to cars, horizontal gaze variability is notably higher than horizontal head movement. This behavior suggests that riders prioritize eye movements for scanning their surroundings while maintaining a relatively steady head position. This approach likely reflects an adaptive strategy to balance stability with hazard monitoring in these high-risk scenarios.

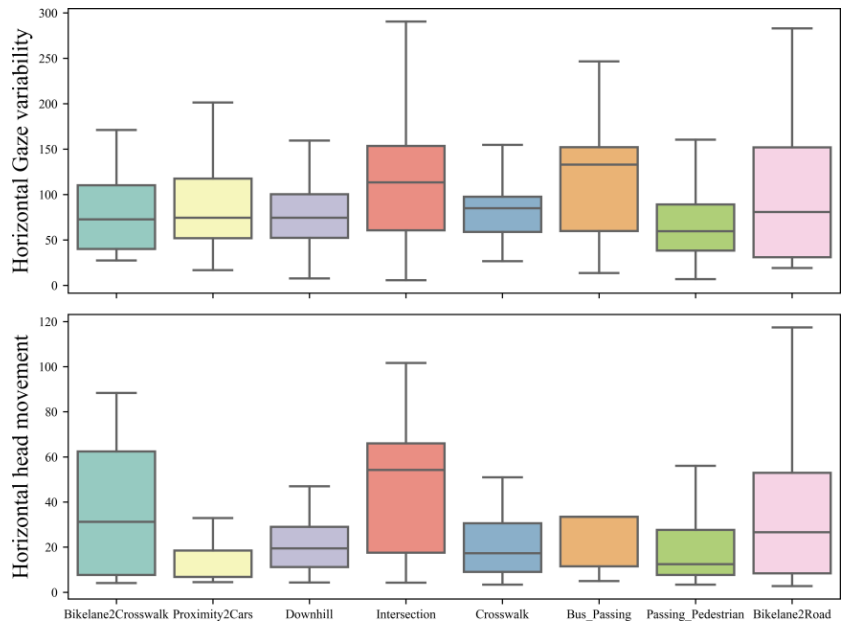

**Fig. 9. Horizontal gaze and head movement variability for different traffic scenarios**

The elevated Stationary Gaze Entropy (SGE) and Gaze Transition Entropy (GTE) observed in scenarios such as downhill riding, intersections, and passing by a bus, as shown in Figure 10, highlight the dynamic and unpredictable visual engagement required of riders. These conditions, characterized by a more challenging and less structured road environment, demand heightened attention and frequent gaze shifts. The higher entropy values indicate that riders must continuously adjust their focus and attention to navigate these scenarios safely. This reflects an increased cognitive load, as riders process and respond to multiple stimuli simultaneously to maintain control and awareness, underscoring the complexity of these specific traffic scenarios.

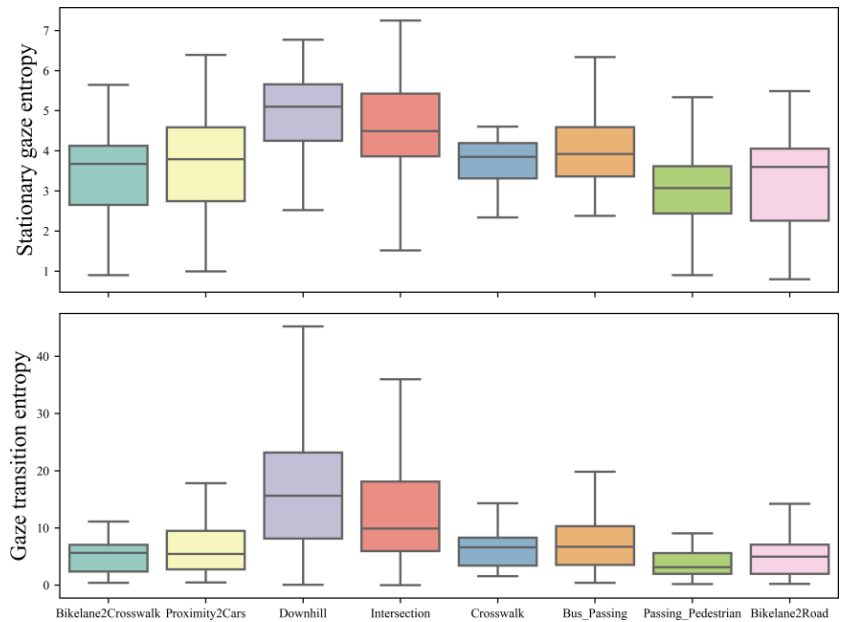

**Fig. 10. Stationary Gaze Entropy (SGE) and Gaze Transition Entropy (GTE) in different traffic scenarios**

## CONCLUSION

This study provides valuable insights into the interplay between road infrastructure, traffic scenarios, and the visual and physical responses of e-scooter riders in a naturalistic setting. The findings highlight how infrastructure types and interactions with traffic influence gaze and head movement patterns, ultimately affecting rider safety and cognitive load. For example, bike lanes were associated with reduced horizontal head movement and a stronger focus on the road surface, suggesting a more predictable riding environment. In contrast, shared roads and sidewalks showed more scattered visual distribution and increased horizontal head movement, indicating a higher level of un- certainty and complexity. When analyzing interactions with other road users, scenarios such as being passed by a bus, riding near cars, navigating intersections, and riding at high speeds on downhill paths were found to demand higher cognitive load. Intersections required heightened visual attention and

Proceedings Paper Formatting Instructions        – 10 –        Rev. 10/2015

spatial awareness, reflected in increased horizontal eye and head movements. In interactions with buses and cars, visual scanning was more active than head movement, likely due to the need to maintain balance and avoid collisions. Meanwhile, high-speed downhill scenarios prompted more focused and centered visual attention to monitor obstacles and road surfaces carefully.

This research aims to offer an initial exploration into gaze behavior under various road and traffic conditions for e-scooter users, rather than drawing definitive conclusions from the limited participant pool. Expanding the participant demographic and skill range in future research is crucial for improving the generalizability of these findings. Additionally, future studies should investigate physiological responses such as heart rate variability, EEG signals, and galvanic skin response to gain a more comprehensive understanding of the cognitive, emotional, and physical challenges faced by e-scooter riders. Such multidimensional analyses could further inform strategies to enhance rider safety and comfort.